\documentclass[runningheads]{llncs}
\usepackage{graphicx}
\usepackage{comment}
\usepackage{amsmath,amssymb} 
\usepackage{xcolor}
\usepackage{epsfig}
\usepackage{soul}
\usepackage[ruled,vlined]{algorithm2e}
\usepackage{enumitem}
\usepackage{booktabs}
\usepackage{background}
\usepackage[T1]{fontenc}

\graphicspath{{./figs/}{./plots}{./images/}}
\newcommand{\eat}[1]{}

\definecolor{stelios_colour}{RGB}{144, 238, 144}

\renewcommand{\st}[1]{}

\newcommand*{\mybox}[1]{\framebox{#1}}

\DeclareMathOperator*{\argmax}{argmax}

\backgroundsetup{angle=0,
    scale=1,
    color=black,
    firstpage=true,
    position=current page.north,
    hshift=0pt,
    vshift=-20pt,
    contents={\ifnum\value{page}=1 PREPRINT: Accepted at the 16th European Conference on Computer Vision (ECCV), 2020 \else \fi}
}

\begin{document}
\pagestyle{headings}
\mainmatter
\def\ECCVSubNumber{5305}  

\title{Journey Towards Tiny Perceptual Super-Resolution} 

\titlerunning{Tiny Perceptual Super-Resolution}
%
\author{Royson Lee\inst{1}
\and
{\L}ukasz Dudziak\inst{1}
\and
Mohamed Abdelfattah\inst{1}
\and
Stylianos I. Venieris\inst{1}
\and
Hyeji Kim\inst{1}
\and
Hongkai Wen\inst{1,2}
\and
Nicholas D. Lane\inst{1,3}
}
\authorrunning{R. Lee et al.}
%
\institute{Samsung AI Center, Cambridge, UK \and
University of Warwick \and 
University of Cambridge\\
\email{\{royson.lee,\,l.dudziak,\,mohamed1.a,\,s.venieris,\\hyeji.kim1,\,hongkai.wen,\,nic.lane\}@samsung.com}}

\maketitle

\begin{abstract}
Recent works in single-image perceptual super-resolution (SR) have demonstrated unprecedented performance in generating realistic textures by means of deep convolutional networks. 
However, these convolutional models are excessively large and expensive, hindering their effective deployment to end devices. 
In this work, we propose a neural architecture search (NAS) approach that integrates NAS and generative adversarial networks (GANs) with recent advances in perceptual SR and pushes the efficiency of small perceptual SR models to facilitate on-device execution.
Specifically, we search over the architectures of both the generator and the discriminator sequentially, highlighting the unique challenges and key observations of searching for an SR-optimized discriminator and comparing them with existing discriminator architectures in the literature.
Our tiny perceptual SR (TPSR) models outperform SRGAN and EnhanceNet on both full-reference perceptual metric (LPIPS) and distortion metric (PSNR) while being up to 26.4$\times$ more memory efficient and 33.6$\times$ more compute efficient respectively.

\end{abstract}
\section{Introduction}
\label{sec:intro}

Single-image super-resolution (SR) is a low-level vision problem that entails the upsampling of a single low-resolution (LR) image to a high-resolution (HR) image. 
Currently, the highest-performing solutions to this problem are dominated by the use of convolutional neural networks, which have left limited space for traditional approaches~\cite{Chang_2004,Kim_2010}. 
Nevertheless, with the super-resolution task being inherently ill-posed, \textit{i.e.} a given LR image can correspond to many HR images, SR methods follow different approaches. In this respect, existing supervised solutions can be mainly grouped into two tracks based on the optimization target: distortion and perceptual quality.

To improve perceptual quality, Ledig \textit{et al.}~\cite{SRGAN} first empirically showed that the use of generative adversarial networks (GANs)~\cite{Goodfellow_2014} results in upsampled images that lie closer to the natural-image manifold. This observation was later backed theoretically~\cite{Blau_2018} through a proof that using GANs is a principled approach to minimize the distance between the distribution of the upsampled image and that of natural images. Until today, there have been several works focusing on using GANs for perceptual SR, leading to prominent networks such as ESRGAN~\cite{ESRGAN} and EnhanceNet~\cite{EnhanceNet}.

Although these proposed perceptual SR solutions achieve promising results, they remain extremely resource-intensive in terms of computational and memory demands. Existing \textit{efficient} SR solutions~\cite{IDN,IMDN,FEQE,FSRCNN,SRCNN,CARN,FALSR,MoreMNAS,ESRN,Lee_2019}, on the other hand, are mostly optimized for distortion metrics, leading to blurry results. 
Hence, in this work, we pose the following question: \textbf{Can we build an efficient and constrained SR model while providing high perceptual quality?}

{In order to build such SR models, we apply neural architecture search (NAS). In particular, we run NAS on both the discriminator as well as the generator architecture.}
To the best of our knowledge, our study is the first to search for a discriminator in SR, shedding light on the role of the discriminator in GAN-based perceptual SR. 
Our contributions can be summarized as follows:
\begin{itemize}
    \item We adopt neural architecture search (NAS) to find efficient GAN-based SR models, using PSNR and LPIPS~\cite{LPIPS} as the rewards for the generator and discriminator searches respectively. 
	\item We extensively investigate the role of the discriminator in training our GAN and we show that both existing and new discriminators of various size and compute can lead to perceptually similar results on standard benchmarks.
	\item We present a tiny perceptual SR (TPSR) model that yields high-performance results in both full-reference perceptual and distortion metrics against much larger full-blown perceptual-driven models.
\end{itemize}

\section{Background \& Related Work}
\label{sec:background}

In SR, there is a fundamental trade-off between distortion- and perceptual-based methods~\cite{Blau_2018}; higher reconstruction accuracy results in a less visually appealing image and vice versa. 
Distortion-based solutions~\cite{SRDenseNet,EDSR,RCAN} aim to improve the fidelity of the upsampled image, \textit{i.e.}~reduce the dissimilarity between the upsampled image and the ground truth, but typically yield overly smooth images. 

Perceptual-based methods~\cite{SRGAN,CX,ESRGAN,EnhanceNet}, on the other hand, aim to improve the visual quality by reducing the distance between the distribution of natural images and that of the upsampled images, resulting in reconstructions that are usually considered more appealing. 
These perceptual SR models are usually commonly evaluated using full-reference methods such as LPIPS~\cite{LPIPS} or no-reference methods such as NIQE~\cite{NIQE}, BRISQUE~\cite{BRISQUE}, and DIIVINE~\cite{DIIVINE}, which are designed to quantify the deviation from natural-looking images in various domains.

\textbf{Hand-crafted Super-resolution Models.}
Since the first CNN was proposed for SR~\cite{SRCNN} there has been a surge of novel methods, adapting successful ideas from other high- and low-level vision tasks. 
For instance, state-of-the-art distortion-driven models such as EDSR~\cite{EDSR}, RDN~\cite{RDN}, and RCAN~\cite{RCAN} use residual blocks~\cite{resnet}, and attention mechanisms~\cite{bahdanau_2015}, respectively, to achieve competitive fidelity results.
Independently, state-of-the-art perceptual-driven SR models have been primarily dominated by GAN-based models such as SRGAN~\cite{SRGAN} (which uses a combination of perceptual loss~\cite{Johnson_2016} and GANs), and ESRGAN~\cite{ESRGAN} (which improves on SRGAN by employing the relativistic discriminator~\cite{JolicoeurMartineau_2018}).

Towards efficiency, Dong \textit{et al.}~\cite{FSRCNN} and Shi \textit{et al.}~\cite{ESPCN} proposed reconstructing the upsampled image at the end of a network, rather than at its beginning, to reduce the computational complexity during feature extraction. 
Since then, numerous architectural changes have been introduced to obtain further efficiency gains. For instance, group convolutions~\cite{resnet} were adopted by Ahn \textit{et al.}~\cite{CARN}, channel splitting~\cite{shufflenetv2} by Hui \textit{et al.}~\cite{IDN,IMDN}, and inverse sub-pixel convolutions by Vu \textit{et al.}~\cite{FEQE}, all of which significantly reduced the computational cost. 

Similar to one of our goals, Chen \textit{et al.}~\cite{chen_2019} explored how the discriminator would affect performance in SR by introducing two types of attention blocks to the discriminator to boost image fidelity in both lightweight and large models. 
Unlike their approach, we optimize for a perceptual metric and explore a wide range of discriminators using standard popular NN operations instead. 

\textbf{Neural Architecture Search for Super-resolution.}
Recent SR works aim to build more efficient models using NAS, which has been vastly successful in a wide range of tasks such as image classification~\cite{zoph_2018,zhong_2018,real_2018}, language modeling~\cite{zoph_2016}, and automatic speech recognition~\cite{dudziak_2019}.
We mainly focus on previous works that adopt NAS for SR and refer the reader to Elsken \textit{et al.}~\cite{elsken_2018} for a detailed survey on NAS. 
Chu \textit{et al.}~\cite{MoreMNAS,FALSR} leveraged both reinforcement learning and evolutionary methods for exploitation and exploration respectively, considering PSNR, FLOPs and memory in a multi-objective optimization problem. 
Song \textit{et al.}~\cite{ESRN} argued that searching for arbitrary combinations of basic operations could be more time-consuming for mobile devices, a guideline that was highlighted by Ma \textit{et al.}~\cite{shufflenetv2}.
To alleviate that, they proposed searching using evolutionary methods for hand-crafted efficient residual blocks.
Although we agree with their approach to utilize platform-specific optimizations, we decided to keep our approach platform-agnostic and only consider models that fit in the practical computational regime based on the models used in the current SoTA SR mobile framework~\cite{Lee_2019}.
Most importantly, our work differs from previous NAS with SR approaches as we focus on optimizing the perceptual quality rather than the fidelity of the upsampled images. 

\textbf{Neural Architecture Search for GANs.}
Recently, Gong \textit{et al.}~\cite{AutoGan} presented a way of incorporating NAS with GANs for image generative tasks, addressing unique challenges faced by this amalgamation. 
Combining NAS with GANs for SR, on the other hand, presents its own set of challenges. 
For example, as perceptual SR only requires one visually appealing solution, mode collapse might be favorable so their proposed dynamic-resetting strategy is not desirable in our context. 
Another major difference is that GAN-based methods for SR usually start with a pre-trained distortion model, avoiding undesired local optima and allowing GAN training to be more stable with high-fidelity input images. Therefore, naively applying their approach is not suitable for the task.
With fewer restrictions, we are able to search for a discriminator as opposed to manually tuning it to fit the generator.

\section{Searching for Tiny Perceptual SR}
\label{sec:approach}

In the proposed scheme, we extend the original REINFORCE-based NAS framework~\cite{zoph_2016} in order to search for a GAN-based super-resolution model. As a first step, we split the process into two stages. 
First, we search only for the best generator, using a selected distortion metric to assess different architectures.
Next, we utilize the best found model and search for a matching discriminator which would maximize the generator's performance on a selected perceptual metric.
Although the same backbone algorithm is used in both cases to conduct the search, the differences between distortion- and GAN-based training require us to approach the two stages with a dedicated methodology, addressing the respective challenges in critical design decisions, including defining the search space and generating reward signals.

\begin{figure}[t]
    \begin{subfigure}{.5\columnwidth}
        \centering
        \includegraphics[scale=0.5]{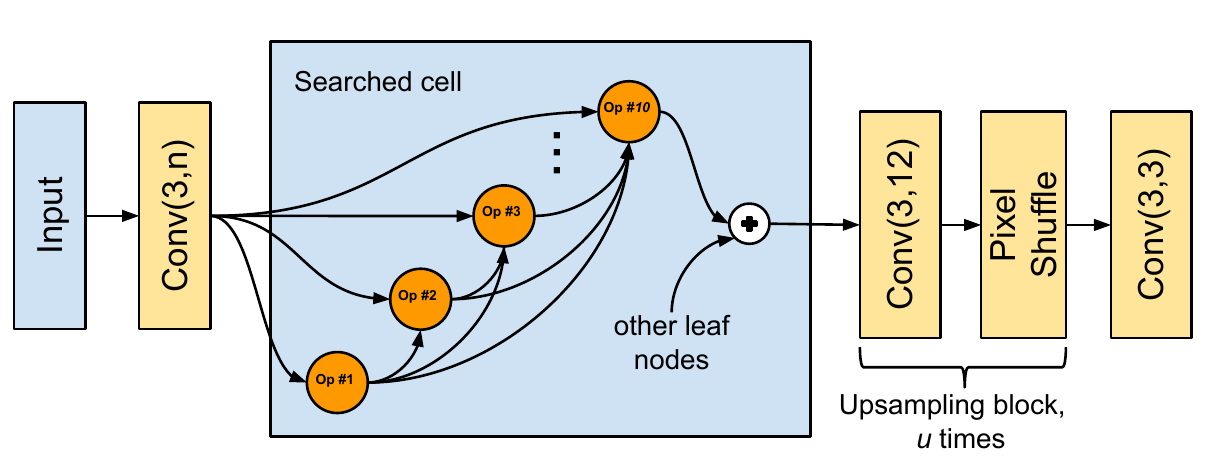}
    \end{subfigure}%
    \begin{subfigure}{.5\columnwidth}
        \centering
        \includegraphics[scale=0.5]{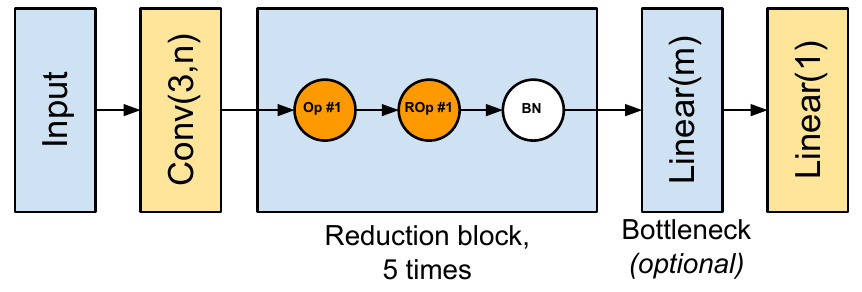}
    \end{subfigure}
    \caption{Structure and search space of the generator (left) and discriminator (right). Orange nodes represent operations which are selected by the controller from the set of available candidates. In the case of the generator, the controller additionally selects only one of the incoming edges as input for each node and, after connections are selected, all leaf nodes are added together to create the cell's output. $\texttt{\small Linear}(n)$ represents a linear layer with $n$ output units. Operations in yellow blocks are fixed.}
    \label{fig:gen_dis_structure}
\end{figure}

We begin with a short introduction to REINFORCE and NAS in Section~\ref{sec:approach:reinforce} and continue to discuss the details related to the specific use-case of perceptual SR.
The skeleton models for both the generator and the discriminator are shown in Figure~\ref{fig:gen_dis_structure} and the search methodology for both of them is presented in Sections~\ref{sec:approach:gen_search} and \ref{sec:approach:dis_search} respectively, with a summary shown in Algorithm~\ref{alg:reinforce}.

\begin{algorithm}[t]
\caption{A summary of the proposed two-stage approach to searching for a perceptually-good compact SR model}
\label{alg:reinforce}
\scriptsize
\LinesNumbered
\SetAlgoLined
\KwIn{search space for the generator $\mathbb{S}_G$ and discriminator $\mathbb{S}_D$, maximum number of steps when searching for generator $T_G$ and discriminator $T_D$, Mult-Adds limit for the generator $f$ }
\KwOut{trained perceptual model $\mathbf{G}^{\bullet}_{\text{best}}$}
\SetKwFunction{FMain}{search}
\SetKwProg{Fn}{Function}{:}{}
\Fn{\FMain{$\mathbb{S}, T, E$}}{
    $s^{\ast}$ $\leftarrow$ \textsc{None} \\
    $\theta  \sim \mathcal{N}$ \\
    \For{$t$ $\leftarrow$ $0$ \KwTo $T$}{
        $s_t \sim \pi_{\theta, \mathbb{S}}$ \\
        $m_t$ $\leftarrow$ $E(\mathbf{s}_t)$ \\
        \If{$m_t=\textsc{None}$}{
            go back to line 5
        }
        update $s^{\ast}$ using $m_t$ \\
        $\theta$ $\leftarrow$ update $\theta$ using $\nabla_{\theta} \log \pi_{\theta,\mathbb{S}}(s_t)R(m_t)$
    }
    \textbf{return} $s^{\ast}$
}
\textbf{End Function} \\[0.2cm]

\SetKwFunction{FMain}{$E_G$}
\SetKwProg{Fn}{Function}{:}{}
\Fn{\FMain{$s$}}{
    $\mathbf{G}$ $\leftarrow$ construct model according to $s$ and initialize its weights with the cached ones\\
    $f_s$ $\leftarrow$ calc Mult-Adds required to run $\mathbf{G}$ \\
    \If{$f_s > f$}{
        \textbf{return} \textsc{None}
    }
    $m$ $\leftarrow$  train and evaluate $\mathbf{G}$ on the proxy distortion task \\
    update cached weights according to Eq.~\ref{eq:sharing_is_caring} \\
    \textbf{return} $m$
}
\textbf{End Function} \\[0.2cm]

$s^{\ast}_{G}$ $\leftarrow$ $\texttt{search}(\mathbb{S}_G, T_G, E_G)$ \\[0.1cm]
$\mathbf{G}_{\text{best}}$ $\leftarrow$ construct model using $s^{\ast}_{G}$, initialize from cache, and train on the full dist. task \\[0.07cm]

\SetKwFunction{FMain}{$E_D$}
\SetKwProg{Fn}{Function}{:}{}
\Fn{\FMain{$\mathbf{s}$}}{
    $\mathbf{D}$ $\leftarrow$ construct discriminator according to $\mathbf{s}$ \\
    \textbf{return} performance of $\mathbf{G}$ on the proxy perc. task after fine-tuning using $\mathbb{D}$
}
\textbf{End Function} \\[0.2cm]

$s^{\ast}_{D}$ $\leftarrow$ $\texttt{search}(\mathbb{S}_D, T_D, E_D)$ \\[0.1cm]
$\mathbf{D}_{\text{best}}$ $\leftarrow$ construct discriminator according to $s^{\ast}_{D}$ \\[0.07cm]
$\mathbf{G}^{\bullet}_{\text{best}}$ $\leftarrow$ fine-tune $\mathbf{G}_{\text{best}}$ with $\mathbf{D}_{\text{best}}$ on the full perceptual task
\end{algorithm}

\subsection{Searching Algorithm}
\label{sec:approach:reinforce}

We can formulate our NAS problem in a generic way as:
\begin{equation}
    \small
    \begin{gathered} 
    \label{eq:opt}
        \mathbb{S} = \mathbb{O}_1 \times \mathbb{O}_2 \times \cdots \times \mathbb{O}_n \\
        E: \mathbb{S} \rightarrow \mathbb{R} \\
        s^{\star} = \argmax_{s\in S} \; E(s)
    \end{gathered}
\end{equation}
where $\mathbb{S}$ is a \textit{search space} constructed from $n$ independent \textit{decisions}, $\mathbb{O}_i$ is a set of available \textit{options} for the $i$-th decision, and $E$ is a selected \textit{evaluation function} which we aim to optimize.

Usually, $E$ is implemented as a sequence of steps: construct a model according to the selected options $s$, train and evaluate it, and return its performance.
In our case specifically, $E$ represents a trained model's performance on a validation set -- see the following sections for the details about training and evaluation of different models.
Because training takes an excessive amount of time and it is hard to predict the performance of a model without it, brute-forcing the optimization problem in Eq.~(\ref{eq:opt}) quickly becomes infeasible as the number of elements in $\mathbb{S}$ increases.
Therefore, a standard approach is to limit the search process to at most $T$ models (steps), where $T$ is usually decided based on the available time and computational resources.
Given a sequence of $T$ architectures explored during the search $\tau(T) = (s_1, s_2, \cdots, s_T)$, we can approximate the optimization problem in Eq.~(\ref{eq:opt}) with its equivalent over the values in $\tau(T)$:
\begin{equation} 
    \begin{gathered} \label{eq:opt_approx}
        s^{\star} \approx s^{\ast} = \argmax_{s \in \tau(T)} \; E(s)
    \end{gathered} 
\end{equation}
We then use REINFORCE~\cite{williams_1992} to guide the search and ensure that, as $T$ increases, $s_T$ optimizes $E$ thus providing us with a better approximation.
More specifically, we include a probabilistic, trainable policy $\pi_\theta$ (a controller) which, at each search step $t=1,2,\cdots,T$, is first sampled in order to obtain a candidate structure $s_t$ and then updated using $E(s_t)$.
We use the following standard formulation to optimize this policy:
\begin{equation} 
    \begin{gathered} \label{eq:reinforce}
        J(\theta) = \mathbb{E}_{s \sim \pi_\theta}R(E(s)) \\
        \theta_{t+1} = \theta_{t} + \beta \nabla_{\theta_t} J(\theta_t) \\
        \nabla_{\theta_t} J(\theta_t) \approx \nabla_{\theta_t} \log \pi_{\theta_t}(s_t)R(E(s_t)) \\
        \argmax_\theta J(\theta) \approx \theta_T
    \end{gathered}
\end{equation}
where $R: \mathbb{R} \rightarrow \mathbb{R}$ is a \textit{reward function} used to make values of $E$ more suitable for the learning process (\textit{e.g.}~via normalization), and $\beta$ is a learning rate. 
Please refer to the supplementary material for further details.

\subsection{Generator Search}
\label{sec:approach:gen_search}

When searching for the generator architecture, we adopt the micro-cell approach. Under this formulation, we focus the search on finding the best architecture of a single cell which is later placed within a fixed template to form a full model.
In conventional works, the full model is constructed by stacking the found cell multiple times, forming in this way a deep architecture. In our case, since we aim to find highly compact models, we defined the full-model architecture to contain a single, relatively powerful cell.
Furthermore, the single cell is instantiated by selecting 10 operations from the set of available candidates, $Op$, to assign to 10 nodes within the cell. The connectivity of each node is determined by configuring the input to its operation, selecting either the cell's input or the output of any previous node. 
Thus, our generator search space (Figure~\ref{fig:gen_dis_structure} (left)) is defined as:
\begin{equation} 
    \begin{gathered} \label{eq:micro_ss}
        \mathbb{S}_G = \mathbb{S}_{\textsc{cell}} = \underbrace{Op \times \mathbb{Z}_{1} \times \cdots \times Op \times \mathbb{Z}_{10}}_{20 \text{ elements}}
    \end{gathered} 
\end{equation}
where $\mathbb{Z}_{m} = \{ 1,2,\cdots, m \}$ is a set of indices representing possible inputs to a node.
We consider the following operations when searching for the generator:
\begin{equation*} 
    \begin{gathered}
    Op = \{\; \texttt{\small Conv}(k,n) \text{ with } k=1,3,5,7\text{; } \texttt{\small Conv}(k,n,4), \\
               \texttt{\small DSep}(k,n), \text{ and } \texttt{\small InvBlock}(k,n,2) \text{ with } k=3,5,7\text{; } \\
               \texttt{\small SEBlock}(), \texttt{\small CABlock}(), \texttt{\small Identity} \;\}
    \end{gathered}
\end{equation*}
where $\texttt{\small Conv}(k,n,g \!\!=\!\! 1, s \!\!=\!\! 1)$ is a convolution with kernel $k\times k$, $n$ output channels, $g$ groups and stride $s$; $\texttt{\small DSep}(k,n)$ is depthwise-separable convolution~\cite{mobilenetv1}; $\texttt{\small SEBlock}$ is Squeeze-and-Excitation block~\cite{SE_block}; $\texttt{\small CABlock}$ is channel attention block~\cite{RCAN}; and $\texttt{\small InvBlock}(k,n,e)$ is inverted bottleneck block~\cite{mobilenetv2} with kernel size $k\times k$, $n$ output channels and expansion of $e$.
In the case where a cell constructed from a point in the search space has more than one node which is not used as input to any other node (\textit{i.e.}~a leaf node), we add their outputs together and use the sum as the cell's output.
Otherwise, the output of the last node is set as the cell's output.

We use weight sharing similar to~\cite{pham_2018}. That is, for each search step $t$, after evaluating an architecture $s_t \in \mathbb{S}_G$ we save trained weights and use them to initialize the weights when training a model at step $t+1$.
Because different operations will most likely require weights of different shape, for each node $i$ we keep track of the best weights so far for each operation from the set $Op$ independently.
Let $s(i)$ be the operation assigned to the $i$-th node according to the cell structure $s$.
Further, let $\mathbb{P}_{o,i,t}$ be the set of architectures explored until step $t$ (inclusive) in which $o$ was assigned to the $i$-th node, that is: $\mathbb{P}_{o,i,t} = \{ s\ |\ s\in \tau(t) \land s(i)=o \}$.
Finally, let $\theta_{i,o,t}$ represent weights in the cache, at the beginning of step $t$, for an operation $o$ when assigned to the $i$-th node, and $\hat{\theta}_{i,o,t}$ represent the same weights after evaluation of $s_t$ (which includes training).
Note that $\hat{\theta}_{o,i,t}=\theta_{o,i,t}$ if $s_t(i)\neq o$ as the weights are not subject to training.
We can then formally define our weight sharing strategy as:
\begin{equation} 
        \begin{gathered} \label{eq:sharing_is_caring}
        \theta_{o,i,0} \sim \mathcal{N} \\[4pt]
        \theta_{o,i,t+1} =
        \begin{cases}
            \hat{\theta}_{o,i,t} & \parbox{.55\columnwidth}{if $s_{t}(i)=o$ and \newline $E(s_t) > \max_{s_p\in \mathbb{P}_{o,i,t-1}}E(s_p)$} \\[6pt]
            \theta_{o,i,t} & \text{otherwise}
        \end{cases}
    \end{gathered}
\end{equation}
As SR models require at least one order of magnitude more compute than classification tasks, we employ a variety of techniques to speed up the training process when evaluating different architectures and effectively explore a larger number of candidate architectures.
First, similar to previous works~\cite{zoph_2018}, we use lower fidelity estimates, such as fewer epochs with higher batch sizes, instead of performing full training until convergence which can be prohibitively time consuming. 
Moreover, we use smaller training patch sizes as previous studies~\cite{ESRGAN} have shown that the performance of the model scales according to its training patch size, preserving in this manner the relative ranking of different architectures. 
Lastly, we leverage the small compute and memory usage of the models in our search space and dynamically assign multiple models to be trained on each GPU.
We also constrain the number of Mult-Adds and discard all proposed architectures which exceed the limit before the training stage, guaranteeing the generation of small models while, indirectly, speeding up their evaluation.

After the search has finished, we take the best found design point $s^{\ast}$ and train it on the full task to obtain the final distortion-based generator $G$, before proceeding to the next stage.
When performing the final training, we initialize the weights with values from the cache $\theta_{o,i,T}$, as we empirically observed that it helps the generator converge to better minima. 
Both the proxy and full task aim to optimize the fidelity of the upsampled image and are, thus, validated using PSNR and trained on the training set $\hat{T}$ using the L1 loss, defined as:

\begin{equation} 
    \label{eq:l1}
    L_1 = \frac{1}{|\hat{T}|} \sum_{(I^{\text{LR}},I^{\text{HR}})\in \hat{T}}|G(I^{\text{LR}}) - I^{\text{HR}}|
\end{equation}
with $I^{\text{LR}}$ the low-resolution image and $I^{\text{HR}}$ the high-resolution ground-truth.

\subsection{Discriminator Search}
\label{sec:approach:dis_search}
After we find the distortion-based generator model $G$, we proceed to search for a matching discriminator $D$ that will be used to optimize the generator towards perceptually-good solutions.
The internal structure of our discriminator consists of 5 reduction blocks.
Each reduction block comprises a sequence of two operations followed by a batch normalization~\cite{BN} -- the first operation is selected from the set of candidate operations $Op$ which is the same as for the generator search; the second one is a reduction operation and its goal is to reduce the spatial dimensions along the x- and y-axes by a factor of 2 while increasing the number of channels by the same factor.
To only choose reduction operations from the set of operations derived from $Op$, we only consider standard convolutions with the same hyperparameters as in $Op$, but with stride changed to 2 and increased number of output channels:
\begin{equation*} 
    \begin{aligned}
        ROp = \{\; & \texttt{\small Conv}(k,2n,1,2) \text{ with } k=1,3,5,7 \text{ and } \\
                   & \texttt{\small Conv}(k,2n,4,2) \text{ with } k=3,5,7 \;\}
    \end{aligned} 
\end{equation*}
As a result, the search space for the discriminator can be defined as:
\begin{equation}
    \mathbb{S}_D = \underbrace{Op \times ROp \times \cdots \times Op \times ROp}_{10 \text{ elements}}
\end{equation}
After the 5 reduction blocks, the extracted features are flattened to a 1-D vector and passed to a final linear layer (preceded by an optional bottleneck with $m$ outputs), producing a single output which is then used to discriminate between the generated upsampled image, $G(I^{\text{LR}})$ and the ground truth, $I^\text{{HR}}$. 
Figure~\ref{fig:gen_dis_structure} (right) shows the overall structure of the discriminator architecture.

To optimize for perceptual quality (Eq.~(\ref{eq:loss})), we use the perceptual loss~\cite{Johnson_2016}, $L_{vgg}$, and adversarial loss~\cite{Goodfellow_2014}, $L_{adv}$, on both the proxy and full task. 
The discriminator is trained on the standard loss, $L_D$. 
As observed by previous works~\cite{ESRGAN,EnhanceNet,chen_2019}, optimizing solely for perceptual quality may lead to undesirable artifacts. 
Hence, similar to Wang \textit{et al.}~\cite{ESRGAN}, we incorporate $L_1$ into the generator loss, $L_G$.
Additionally, we validate the training using a full-reference perceptual metric, Learned Perceptual Image Patch Similarity~\cite{LPIPS} (LPIPS), as we find no-reference metrics such as NIQE to be more unstable since they do not take into account the ground truth.
Our generator loss and discriminator loss are as follows:

\begin{equation} 
    \begin{gathered}
        L_{vgg} = \frac{1}{|\hat{T}|} \sum_{(I^{LR},I^{HR})\in \hat{T}}(\phi(G(I^{\text{LR}})) - \phi(I^{\text{HR}}))^2 \\
        L_{adv} = -\log(D(G(I^{\text{LR}})) \\
        L_G = \alpha L_{1} + \lambda L_{vgg} + \gamma L_{adv} \\
        L_D = -\log(D(I^{\text{HR}})) - \log(1 - D(G(I^{\text{LR}}))) \\
        \label{eq:loss}
    \end{gathered}
\end{equation}
Unlike the generator search we do not use weight sharing when searching for the discriminator.
The reason behind this is that we do not want the discriminator to be too good at the beginning of training to avoid a potential situation where the generator is unable to learn anything because of the disproportion between its own and a discriminator's performance.
Similar to the generator search stage, we used a lower patch size, fewer epochs, and a bigger batch size to speed up the training. 
Additionally, from empirical observations in previous works~\cite{Johnson_2016,SRGAN}, the perceptual quality of the upsampled image scales accordingly with the depth of the network layer used. 
Therefore, we use an earlier layer of the pre-trained VGG network ($\phi$) as a fidelity estimate to save additional computations.
In contrast to the generator search, we do not impose a Mult-Adds limit to the discriminator as its computational cost is a secondary objective for us, since it does not affect the inference latency upon deployment.

After the search finishes, we collect a set of promising discriminator architectures and use them to train the generator on the full task.
At the beginning of training, we initialize the generator with the pre-trained model found in the first stage (Section~\ref{sec:approach:gen_search}) to reduce artifacts and produce better visual results.

\section{Evaluation}
\label{sec:eval}

In this section, we present the effectiveness of the proposed methodology. 
For all experiments, the target models were trained and validated on the DIV2K~\cite{DIV2K} dataset and tested on the commonly-used SR benchmarks, namely Set5~\cite{Set5}, Set14~\cite{Set14}, B100~\cite{B100}, and Urban100~\cite{Urban100}. 
For distortion (PSNR/Structural Similarity index (SSIM)~\cite{SSIM}) and perceptual metrics (Natural Image Quality Evaluator~\cite{NIQE} (NIQE)/Perceptual Index~\cite{Blau_2018b} (PI)), we shaved the upsampled image by its scaling factor before evaluation. 
For LPIPS, we passed the whole upsampled image and evaluated it on version 0.1 with linear calibration on top of intermediate features in the VGG~\cite{VGG} network\footnote{Provided by https://github.com/richzhang/PerceptualSimilarity}. For the exhaustive list of all hyperparameters and system details, please refer to the supplementary material.

\subsection{TPSR Generator}

Following Algorithm~\ref{alg:reinforce}, we began by running the first search stage for the generator architecture to obtain a distortion-driven tiny SR model.
During the search, we trained candidate models to perform $\times 2$ upscaling (\textit{i.e.} with one upsampling block) and we set the number of feature maps ($n$) to 16.
Each model was evaluated using PSNR as the target metric and the final reward for the controller was calculated by normalizing the average PSNR of a model.

\begin{figure}[t]
    \centering
    {
    \includegraphics[width=0.35\columnwidth,trim=1cm 0cm 1.9cm 0cm,clip,angle=90]{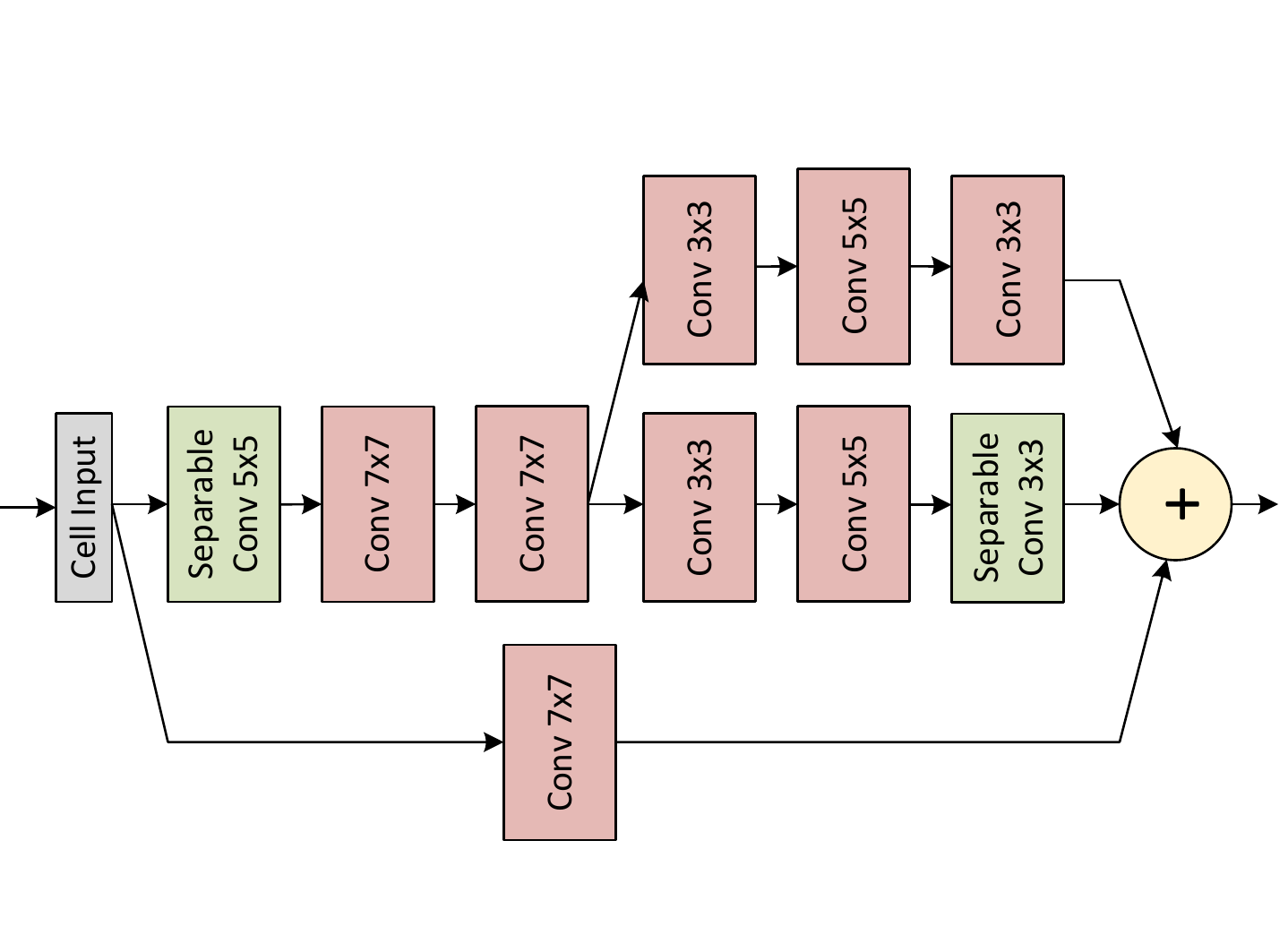}
    }
    \caption{Discovered cell architecture for the TPSR generator. Each operation is followed by a PReLU~\cite{prelu} activation.}
    \label{fig:gen_cell}
\end{figure}

To obtain the final generator model, we run the generator search for 2,500 steps and stored the highest-performing cell architecture as evaluated on the proxy task. 
Figure~\ref{fig:gen_cell} illustrates the obtained cell structure. We refer to this model as TPSR (Tiny Perceptual Super Resolution). For the rest of this section, we use the notation TPSR-$X$ to refer to TPSR when trained with discriminator $X$ and TPSR-NOGAN when TPSR is distortion-driven.

After the end of the first search stage, we trained the discovered TPSR model on the full task for $\times$2 upscaling and $\times$4 upscaling, starting from the pre-trained $\times$2 model, to obtain TPSR-NOGAN. 
Our NAS-based methodology was able to yield the most efficient architecture of only 3.6G Mult-Adds on $\times$4 upscaling with performance that is comparable with the existing state-of-the-art distortion-driven models that lie within the same computational regime. 
Given that our goal was to build a perceptual-based model, we did not optimize our base model further, considering it to be a good basis for the subsequent search for a discriminator. 
The distortion-based results can be found in the supplementary material.

\subsection{Discriminator Analysis}
\label{sec:eval:dis}

\begin{figure}[t]
    \centering
    \includegraphics[scale=0.4]{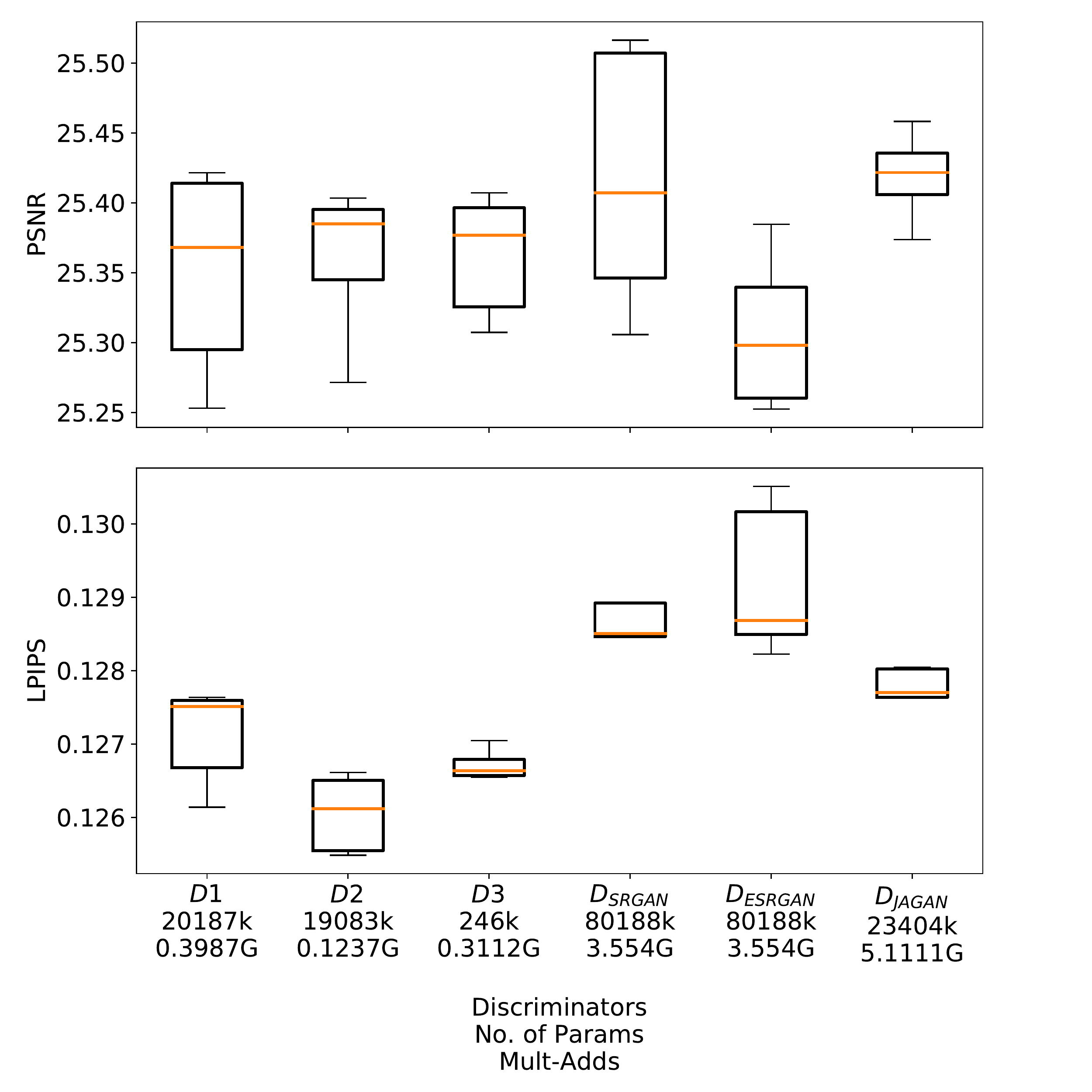}
    \caption{
    Performance of TPSR after adversarial training using different discriminators found via NAS (D1, D2, D3) vs. existing discriminators designed for (SRGAN, ESRGAN, JAGAN). TPSR trained on searched discriminators, which are optimized for LPIPS, outperform fixed discriminators in the literature for the targeted metric (LPIPS). Each GAN training was performed 5 times}
    \label{fig:d_search}
\end{figure}

To obtain a discriminator architecture, we utilized the TPSR-NOGAN variant trained on the $\times$4 upscaling task and searched for a suitable discriminator to minimize the perceptual LPIPS metric. 
To minimize the instability of the perceptual metric, we evaluated each model by considering the last three epochs and returning the best as the reward for the controller.
We also incorporated spectral normalization~\cite{spectral_norm} for the discriminator on both the proxy and the full task.
We have found that discriminators of varying size and compute can lead to perceptually similar results (LPIPS).
Upon further examinations, we have also found that these upsampled images look perceptually sharper than TPSR-NOGAN's.

\setlength{\tabcolsep}{3pt}
\begin{table*}[t]
\begin{center}
\caption{We compare our $\times$4 upscaling TPSR models, which are optimized for LPIPS, with perceptual-driven models in the literature. Higher is better for PSNR and lower is better for LPIPS and PI. \textcolor{red}{red}/\textcolor{blue}{blue} represents \textcolor{red}{best}/\textcolor{blue}{second best} respectively. On the optimization target metric, LPIPS, our model (TPSR-D2) achieves the \emph{second best} result while it is the \emph{smallest} among all. Our model outperforms EnhanceNet and SRGAN in visual quality metrics (PSNR \& LPIPS) while being 26.4$\times$ more memory efficient and 33.6$\times$ more compute efficient than SRGAN and EnhanceNet, respectively}
\resizebox{0.98\textwidth}{!}{
\begin{tabular}{lrrllll}
\toprule
\multicolumn{1}{c}{Model} & \multicolumn{1}{c}{\begin{tabular}[c]{@{}c@{}}Params\\ (K)\end{tabular}} & \multicolumn{1}{c}{\begin{tabular}[c]{@{}c@{}}Mult-Adds\\ (G)\end{tabular}} & \multicolumn{1}{c}{\begin{tabular}[c]{@{}c@{}}Set5\\ PSNR/LPIPS/PI\end{tabular}} & \multicolumn{1}{c}{\begin{tabular}[c]{@{}c@{}}Set14\\ PSNR/LPIPS/PI\end{tabular}} & \multicolumn{1}{c}{\begin{tabular}[c]{@{}c@{}}B100\\ PSNR/LPIPS/PI\end{tabular}} & \multicolumn{1}{c}{\begin{tabular}[c]{@{}c@{}}Urban100\\ PSNR/LPIPS/PI\end{tabular}} \\ \midrule
ESRGAN & 16,697 & 1034.1 & \textcolor{blue}{30.40}/\mybox{\textcolor{red}{0.0745}}/3.755 & 26.17/\mybox{\textcolor{red}{0.1074}}/\textcolor{blue}{2.926} & 25.34/\mybox{\textcolor{red}{0.1083}}/\textcolor{blue}{2.478} & \textcolor{blue}{24.36}/\mybox{\textcolor{red}{0.1082}}/3.770 \\
SRGAN & 1,513 & 113.2 & 29.40/0.0878/\textcolor{blue}{3.355} & 26.05/0.1168/\textcolor{red}{2.881} & 25.19/0.1224/\textcolor{red}{2.351} & 23.67/0.1653/\textcolor{red}{3.323} \\
EnhanceNet & 852 & 121.0 & 28.51/0.1039/\textcolor{red}{2.926} & 25.68/0.1305/3.017 & 24.95/0.1291/2.907 & 23.55/0.1513/\textcolor{blue}{3.471} \\
FEQE & \textcolor{blue}{96} & \textcolor{blue}{5.64} & \textcolor{red}{31.29}/0.0912/5.935 & \textcolor{red}{27.98}/0.1429/5.400 & \textcolor{red}{27.25}/0.1455/5.636 & \textcolor{red}{25.26}/0.1503/5.499 \\
TPSR-D2 & \textcolor{red}{61} & \textcolor{red}{3.6} & 29.60/\mybox{\textcolor{blue}{0.076}}/4.454 & \textcolor{blue}{26.88}/\mybox{\textcolor{blue}{0.110}}/4.055 & \textcolor{blue}{26.23}/\mybox{\textcolor{blue}{0.116}}/3.680 & 24.12/\mybox{\textcolor{blue}{0.141}}/4.516 \\
\bottomrule
\end{tabular}}
\label{perceptual_table}
\end{center}
\end{table*}

In order to evaluate the fidelity of the proxy task, we took the three best performing discriminator candidates based on their performance on the proxy task.
We then evaluated our TPSR model when trained with these discriminators on the full task.
To compare to models from the literature, we also considered the discriminators that were used in SRGAN~\cite{SRGAN}, ESRGAN~\cite{ESRGAN}, and the recently proposed Joint-Attention GAN~\cite{chen_2019} (JAGAN).
Note that the discriminator's architecture in SRGAN and ESRGAN is the same but the latter is trained using the relativistic GAN (RGAN) loss~\cite{JolicoeurMartineau_2018}.
For more details on how RGAN is adopted for SR, please refer to Wang \textit{et al.}~\cite{ESRGAN}.

\begin{figure}[t]
    \centering
    \includegraphics[width=0.525\columnwidth,trim=0 0cm 0 0.1cm,clip]{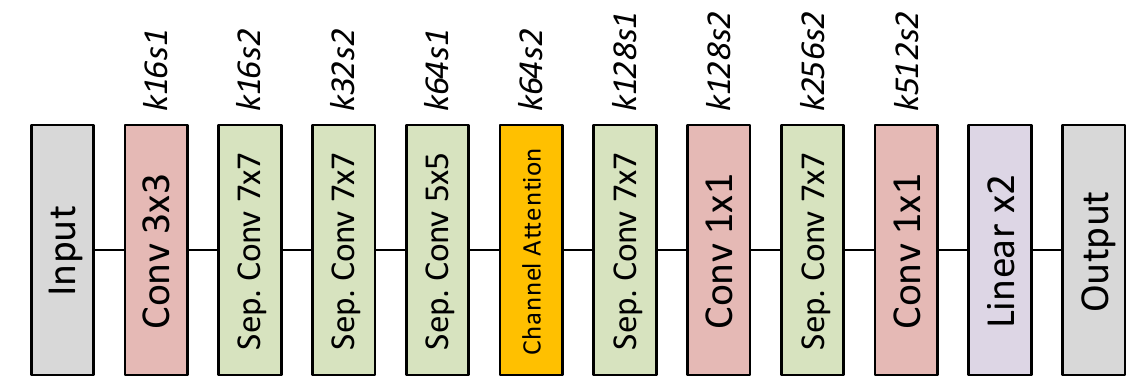}
    \caption{Discovered architecture of the TPSR discriminator ($D2$). Convolutions are followed by batch normalization and PReLU. k = output tensor depth, s = stride}
    \label{fig:dis_structures}
\end{figure}

Each GAN training was performed 5 times and the best performing model, based on the achieved LPIPS on the validation set, was evaluated on the test benchmarks. 
Specifically, we took the weighted average (based on the number of images) over the test benchmarks of three metrics: PSNR, NIQE, and LPIPS, and present our findings in Figure~\ref{fig:d_search}. Our chosen discriminators (TPSR-D1, TPSR-D2, TPSR-D3) have led to better results as compared to the existing discriminators ($\text{TPSR-D}_{SRGAN}$, $\text{TPSR-D}_{ESRGAN}$, $\text{TPSR-D}_{JAGAN}$) in the particular perceptual metric (LPIPS) that they were optimized for.
The discriminator of $\text{TPSR-D2}$ can be found in Figure~\ref{fig:dis_structures}.

\begin{figure*}[t]
    \begin{subfigure}{\textwidth}
        \centering
        \includegraphics[width=.75\textwidth]{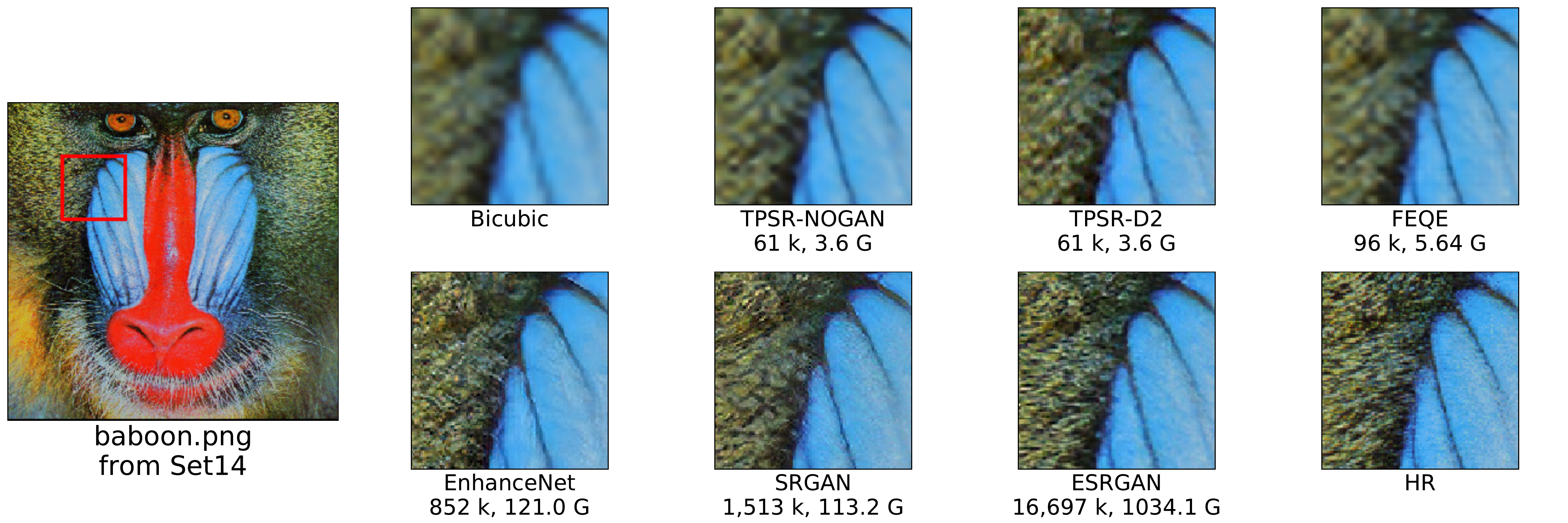}
        \label{fig:vis_comparison:baboon}
    \end{subfigure}
    \begin{subfigure}{\textwidth}
        \centering
        \includegraphics[width=.75\textwidth]{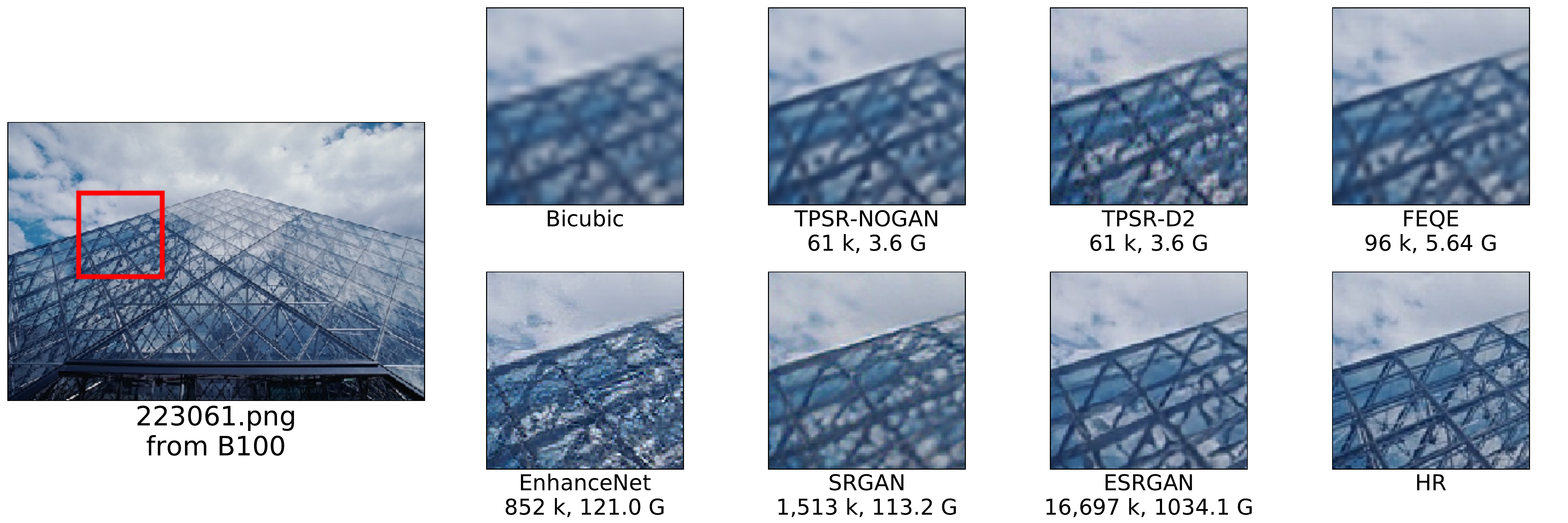}
        \label{fig:vis_comparison:building}
    \end{subfigure}
    \begin{subfigure}{\textwidth}
        \centering
        \includegraphics[width=.75\textwidth]{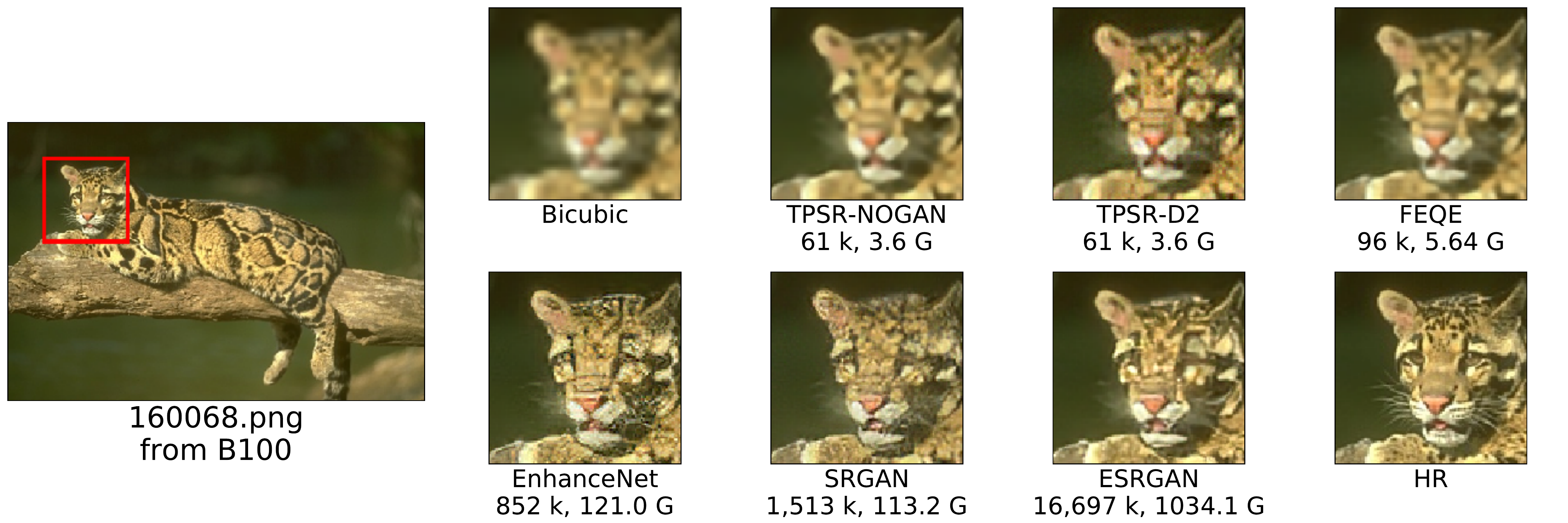}
        \label{fig:vis_comparison:leopard}
    \end{subfigure}
    \caption{Visual comparisons among SoTA perceptual-driven networks and TPSR models, with their no. of parameters (left) and mult-add operations (right). Despite the quantitative results that show that TPSR-D2 is better than eg. SRGAN (Table.~\ref{perceptual_table}), the qualitative results are arguably worse-off in some images, highlighting a limitation and the need for a better perceptual metric. However, TPSR-D2 still produces better reconstructions than FEQE - the current SoTA for constrained perceptual models.}
    \label{fig:vis_comparison}
\end{figure*}

Finally, we compared the best performing GAN-based generator (TPSR-D2) on common full-reference and no-reference perceptual metrics with various well-known perceptual models (Table \ref{perceptual_table}).  
Considering our optimized metric, our model outperforms SRGAN and EnhanceNet while being up to 26.4$\times$ more memory efficient when compared to EnhanceNet and 33.6$\times$ more compute efficient compared to SRGAN. 
Additionally, our model also achieves higher performance in distortion metrics, indicating higher image fidelity and, therefore, constitutes a dominant solution for full-reference metrics (PSNR \& LPIPS) especially with our tiny computational budget. 
Visual comparisons can be found in Figure~\ref{fig:vis_comparison}.

\section{Limitations and Discussion}
\label{sec:limitation}

In this paper, we have presented a NAS-driven framework for generating GAN-based SR models that combine high perceptual quality with limited resource requirements. Despite introducing the unique challenges of our target problem and showcasing the effectiveness of the proposed approach by finding high-performing tiny perceptual SR models, we are still faced with a few open challenges.

The usefulness of NAS approaches which utilize a proxy task to obtain feedback on candidate architectures naturally depends on the faithfulness of the proxy task with regards to the full task.
As GANs are known to be unstable and hard to train~\cite{Salimans_2016}, providing the search with a representative proxy task is even more challenging for them than for other workloads.
We were able to partially mitigate this instability by smoothing out accuracy of a trained network, as mentioned in Section~\ref{sec:eval:dis}. Nevertheless, we still observed that the informativeness of results obtained on the proxy task for GAN-training is visibly worse than \textit{e.g.} results on the proxy task when searching for a generator in the first phase of our method.
This instability is reinforced even more when using no-reference perceptual metrics such as NIQE~\cite{NIQE} and PI~\cite{Blau_2018b} - in which case we observed that training a single model multiple times on our proxy task can result in a set of final accuracies with variance close to the variance of all accuracies of all models explored during the search - rendering it close to useless in the context of searching.
In this respect, we adopted LPIPS which, being a full-reference metric, was able to provide the search with a more robust evaluation of proposed architectures.
While the strategies we used to improve the stability of the search were adequate for us to obtain decent-performing models, the challenge still remains open and we strongly suspect that overcoming it would be a main step towards improving the quality of NAS with GAN-based perceptual training.

Another important challenge comprises the selection of a metric that adequately captures perceptual quality. Identifying a metric that closely aligns with human-opinion scores across a wide range of images still constitutes an open research problem with significant invested research effort \cite{DIIVINE,NIQE,LPIPS,Blau_2018b}. In this respect, although we show that optimizing for LPIPS on the average leads to better quantitative results, the inherent limitations of the metric itself might not result to qualitatively better results on certain images. 

With our work targeting highly compact models optimized for perceptual quality, it is currently challenging to find appropriate baselines that lie within the same computational and memory footprint regime, as FEQE~\cite{FEQE} is, to the best of our knowledge, the only perceptual SR model that meets these specifications.
As a result, in this paper, we also present comparisons with significantly larger models, including SRGAN and EnhanceNet, which our method outperforms in our optimized metric. We also compare with ESRGAN which is more than an order of magnitude more expensive than all examined models. Although our design did not outperform ESRGAN, we could extend our method to explore relaxed constraints to allow a slightly larger generator and employ a relativistic discriminator \cite{JolicoeurMartineau_2018}.
As our focus is on pushing the limits of building a constrained and perceptual SR model that can be deployed in a mobile SR framework~\cite{Lee_2019}, we leave the trade-off between model size and perceptual quality as future work.

Lastly, our method resulted in discriminators that are slightly better than existing discriminators in terms of perceptual performance. 
Nevertheless, even though the performance gains on LPIPS are marginal, our resulted discriminators are orders of magnitude smaller in terms of model size and computational cost and the obtained gains are consistently better across multiple runs. 


\section{Conclusion}
\label{sec:conclusion}

In this paper, we investigated the role of the discriminator in GAN-based SR and the limits to which we can push perceptual quality when targeting extremely constrained deployment scenarios. 
In this context, we adopted the use of NAS to extensively explore a wide range of discriminators, making the following key observations on NAS for GAN-based SR: 
1) Discriminators of drastically varying sizes and compute can lead to similar perceptually good images; possible solutions for the ill-posed super-resolution problem.
2) Due to this phenomenon and the high variance in the results of popular perceptual metrics, designing a faithful proxy task for NAS is extremely challenging. 
Nevertheless, we are able to find discriminators that are consistently better than existing discriminators on our chosen metric, generating a tiny perceptual model that outperforms the state-of-the-art SRGAN and EnhanceNet in both full-reference perceptual and distortion metrics with substantially lower memory and compute requirements. 

\clearpage
%
%
\bibliographystyle{splncs04}
\bibliography{eccv2020submission}

\newpage
\appendix
\section*{Experiment Details}
\label{sec:supp}

\textbf{Neural Architecture Search:}
Similar to previous approaches, we use a single LSTM layer with 100 hidden units as the trainable policy, $\pi_\theta$.
It takes an empty embedding as input and generates a sequence of $l$ outputs, where $l$ is the number of decisions to make in order to decide about a generator's structure.
Each element of the sequence at position $i$ is then passed through the following composition of functions, including a tanh constant of $2.5$ and a sampling logit temperature of $5.0$, in order to reduce its dimensionality and produce a probability distribution:
\begin{equation}
\label{prob_dist}
\text{softmax} \circ 2.5 \cdot \text{tanh} \circ 0.2 \cdot \text{linear} \circ l_i
\end{equation}

Each search is ran on five servers, totalling to 40 NVIDIA GeForce GTX 1080 Ti, and each GPU can fit up to three models which are trained in parallel depending on the sampled model's memory usage at each step. 
Our generator search took 2 days and our discriminator search took 10 days due to the memory requirement needed for the pre-trained generator, sampled discriminator, and two VGG networks to compute $L_{vgg}$ and LPIPS respectively in each training pipeline.
Due to the huge resource needed to run discriminator search, we limit the number of mult-add operations and run a constrained search for a generator, as mentioned in the paper. 
The performance of our generator for the discriminator search is listed in Table ~\ref{distortion_table}.

\begin{table*}[htb]
\begin{center}
\caption{Our TPSR-NOGAN model \textbf{serves as a good basis} for the $\times$4 upscaling discriminator search as it is the most computationally efficient and has performance that is comparable with other distortion-driven models in the literature within the same computational regime. Given that our goal is to build a \textbf{perceptual-based} model, we do not optimize our base model further. Higher is better for distortion metrics. \textcolor{red}{red}/\textcolor{blue}{blue} represents \textcolor{red}{best}/\textcolor{blue}{second best} respectively }
\resizebox{0.9\textwidth}{!}{
\begin{tabular}{llrrllll}
\hline
Scale & \multicolumn{1}{c}{Model} & \multicolumn{1}{c}{\begin{tabular}[c]{@{}c@{}}Params\\ (K)\end{tabular}} & \multicolumn{1}{c}{\begin{tabular}[c]{@{}c@{}}Mult-Adds\\ (G)\end{tabular}} & \multicolumn{1}{c}{\begin{tabular}[c]{@{}c@{}}Set5\\ PSNR/SSIM\end{tabular}} & \multicolumn{1}{c}{\begin{tabular}[c]{@{}c@{}}Set14\\ PSNR/SSIM\end{tabular}} & \multicolumn{1}{c}{\begin{tabular}[c]{@{}c@{}}B100\\ PSNR/SSIM\end{tabular}} & \multicolumn{1}{c}{\begin{tabular}[c]{@{}c@{}}Urban100\\ PSNR/SSIM\end{tabular}} \\ \hline
\multicolumn{1}{c}{} & FSRCNN~\cite{FSRCNN} & \textcolor{red}{12} & \textcolor{blue}{6.0} & 37.00/0.9558 & 32.63/0.9088 & \textcolor{blue}{31.53}/\textcolor{blue}{0.8920} & 29.88/0.9020 \\   
\multicolumn{1}{c}{$\times$2} & MOREMNAS-C~\cite{MoreMNAS} & \textcolor{blue}{25} & \textcolor{red}{5.5} & \textcolor{blue}{37.06}/\textcolor{blue}{0.9561} & \textcolor{blue}{32.75}/\textcolor{blue}{0.9094} & 31.50/0.8904 & \textcolor{blue}{29.92}/\textcolor{blue}{0.9023} \\  
\multicolumn{1}{c}{} & TPSR-NOGAN & 60 & 14.0 & \textcolor{red}{37.38}/\textcolor{red}{0.9583} & \textcolor{red}{33.00}/\textcolor{red}{0.9123} & \textcolor{red}{31.75}/\textcolor{red}{0.8942} & \textcolor{red}{30.61}/\textcolor{red}{0.9119} \\
\hline
 & FSRCNN~\cite{FSRCNN} & \textcolor{red}{12} & \textcolor{blue}{4.6} & 30.71/0.8657 & 27.59/0.7535 & 26.98/0.7150 & 24.62/0.7280 \\   
 \multicolumn{1}{c}{$\times$4} & FEQE-P~\cite{FEQE} & 96 & 5.6 & \textcolor{red}{31.53}/\textcolor{red}{0.8824} & \textcolor{red}{28.21}/\textcolor{red}{0.7714} & \textcolor{red}{27.32}/\textcolor{red}{0.7273} & \textcolor{red}{25.32}/\textcolor{red}{0.7583} \\ 
 & TPSR-NOGAN & \textcolor{blue}{61} & \textcolor{red}{3.6} & \textcolor{blue}{31.10}/\textcolor{blue}{0.8779} & \textcolor{blue}{27.95}/\textcolor{blue}{0.7663} & \textcolor{blue}{27.15}/\textcolor{blue}{0.7214} & \textcolor{blue}{24.97}/\textcolor{blue}{0.7456} \\
 \hline
\end{tabular}
}

\label{distortion_table}
\end{center}
\end{table*}

Each training pipeline is run by a separate process/worker and each worker asynchronously samples the probability distribution (Eq (\ref{prob_dist})), trains the resulting model, and returns the result (PSNR for generator search and LPIPS for discriminator search) from the validation set, $\hat{V}$, of the proxy task. 
The result would then be normalized using a minmax normalization, $N$, to scale it from $0$ to $1$ with a exponential moving average baseline, $EMA$, (decay of 0.95) to obtain the reward. 
We then add the sampled entropy, $H$, ($ \zeta=0.0001$) to the reward and use it to update $\pi_\theta$ via REINFORCE using an Adam optimizer ($\beta_1 = 0.9, \beta_2 = 0.999, \epsilon = 10^{-8}$) with learning rate of $3.5e-4$. 
The training process is detailed in Eq.~\ref{detailed_eq} where $\tilde{M}$ is the metric of choice (PSNR, LPIPS, etc), and $\tilde{L}$ is the training loss ($L_G$ and $L_D$ for generator and discriminator search respectively).

\begin{equation}
\begin{gathered}
    \label{detailed_eq}
R(s)=EMA(N(E(s)))+\zeta H(\pi) \\
E(s)=\frac{1}{|\hat{V}|}\sum_{(I^{LR},I^{HR})\in \hat{V}}  \tilde{M}(G_s(I^{LR},W^{\ast}_s), I^{HR}) \\
W^{\ast}_s=\arg\min_{W_s} \frac{1}{|\hat{T}|}\sum_{(I^{LR},I^{HR})\in \hat{T}} \tilde{L}(G_s(I^{LR},W_s),I^{HR})
\end{gathered}
\end{equation}

\noindent \textbf{Super-resolution:}
The set of hyper-parameters chosen for both the proxy task and the full task is summarized in Table~\ref{hyperparams_gen}.

\begin{table}[htb]
    \caption{Hyper-parameters for the searching (proxy-task) and training (full-task) of the generator and discriminator. The generator search is done on $\times$2 upscaling and the discriminator is done on $\times$4 upscaling. We use the features before activations in the pre-trained VGG19 network provided by PyTorch to compute $L_{vgg}$. Each input patch is an RGB image. We used the same model in both proxy and full task to closely align the performance between both tasks. For speed ups during the search, we use lower fidelity estimates, lower patch size etc, that have been shown in previous works such as ESRGAN to preserve the ranking of the model.}
\begin{center}

\small
\begin{tabular}{l || c | c c}
    Search & Hyper-parameter & Proxy-task & Full-task \\ \hline
    Generator & Epochs & 50 & 450  \\ 
    & Batch size & 64 & 16 \\ 
    & Input patch size & 12$\times$12 & 96$\times$96 \\
    & $L_G$ ($\alpha$) & $1$ & $1$ \\ \hline \hline
    Discriminator & Epochs & 50 & 450  \\ 
    & Batch size & 32 & 16 \\ 
    & Input patch size & 24$\times$24 & 48$\times$48\\ 
    & VGG19 Features & 22 & 54 \\ 
    & $L_G$ ($\alpha$, $\lambda$, $\gamma$) & ($0.01$, $1$, $0.005$) & ($0.01$, $1$, $0.005$) 

\end{tabular}
\label{hyperparams_gen}
\end{center}
\end{table}

For all experiments, we use 800 images for training and 100 images for validation from the DIV2K dataset. 
We train for the stated number of epochs in Table~\ref{hyperparams_gen} using an Adam optimizer ($\beta_1 = 0.9, \beta_2 = 0.999, \epsilon = 10^{-8}$) with learning rate of $1e-4$ at the beginning and $5e-5$ after 200 epochs for the full task. All training patches are randomly flipped, both horizontally and vertically, rotated $90^{\circ}$, and subtracted by the mean RGB values of the DIV2K dataset. 
All operations in the generator is followed by a PReLU and all operations in the discriminator is followed by batch normalization and PReLU.

Finally, all experiments are built on top of EDSR's~\cite{EDSR} code base\footnote{https://github.com/thstkdgus35/EDSR-PyTorch} using PyTorch 1.2. PSNR and SSIM~\cite{SSIM} were evaluated on each image's Y-channel and NIQE~\cite{NIQE} and PI~\cite{Blau_2018b} were evaluated using the official code for the PIRM 2018 SR Challenge\footnote{https://github.com/roimehrez/PIRM2018} on Matlab R2018b. 
As mentioned in the main paper, all images are shaved by their scaling factor before evaluation apart from that of LPIPS (version 0.1), which we evaluate using the linear calibration of the features in the provided VGG network\footnote{https://github.com/richzhang/PerceptualSimilarity}. 
In order to compare to previous works, the number of Mult-Add operations are calculated by upscaling to an 1280 * 720 image.
Images from state-of-the-art models are taken from Wang \textit{et al.}~\cite{ESRGAN}\footnote{https://github.com/xinntao/ESRGAN} and Thang \textit{et al.}~\cite{FEQE}\footnote{https://github.com/thangvubk/FEQE} or generated using provided model by Dong \textit{et al.}~\cite{SRGAN}\footnote{https://github.com/tensorlayer/srgan}.

\end{document}